\documentclass[12pt]{article}

\def\ha{H$\alpha$}                        
\def\habf{H$\mathbf{\alpha}$}    
\def\j13{\textit{Swift}\,J1357.2$-$0933}  
\def\degr{\mbox{$^{\circ}$}}              
\def\msun{\,M$_{\odot}$}                  
\def\rsun{\,R$_{\odot}$}                  
\def\kms{\,km\,s$^{-1}$}                  
\def\up#1{\hbox{${}^{\hbox{\sc #1}}$}}   

\usepackage{amssymb}             
\usepackage{caption}            
\usepackage{url}                
\usepackage{epsfig,graphicx}
\usepackage{rotating}
\usepackage{color}
\usepackage{ifpdf}

\ifpdf
  \usepackage[pdftex, pdfborder=0, breaklinks, baseurl=http://, pdfpagemode=None, pdfstartview=XYZ, pdfstartpage=1, colorlinks=true, linkcolor=blue, citecolor=black, urlcolor=blue]{hyperref}
  \hypersetup{
}
\else
  \usepackage[dvips]{hyperref}
\fi

\ifpdf
  \usepackage{pdfpages}
\fi

\usepackage{scicite}

\usepackage{times}

\newenvironment{sciabstract}{%
\begin{quote} \bf}
{\end{quote}}

\newcounter{lastnote}

\title{A Black Hole Nova Obscured by an Inner Disk Torus}

\author
{J.~M.~Corral-Santana,$^{\ast\,1,2}$ J. Casares,$^{1,2}$  
T.~Mu\~noz-Darias,$^{3}$ P.~Rodr\'iguez-Gil,$^{1,2}$\\
T.~Shahbaz,$^{1,2}$ M.~A.~P.~Torres,$^{4,5}$ 
C.~Zurita,$^{1,2}$ A.~A.~Tyndall,$^{6,7,8}$\\
\\
\normalsize{\up{{1}} Instituto de Astrof\'isica de Canarias,}\\
\normalsize{E--38205 La Laguna, S/C de Tenerife, Spain}
\\
\normalsize{\up{{2}} Departamento de Astrof\'isica, Universidad de La Laguna}\\
\normalsize{ E--38206 La Laguna, S/C de Tenerife, Spain}
\\
\normalsize{\up{{3}} School of Physics and Astronomy, University of Southampton, }\\
\normalsize{ Southampton, SO17 1BJ, United Kingdom}
\\
\normalsize{\up{{4}} SRON, Netherlands Institute for Space Research,}\\
\normalsize{ 3584 CA, Utrecht, The Netherlands}
\\
\normalsize{\up{{5}} Harvard-Smithsonian Center for Astrophysics,}\\
\normalsize{ 60 Garden Street, Cambridge, MA 02138, USA}
\\
\normalsize{\up{{6}} Jodrell Bank Centre for Astrophysics, Univ. of Manchester}\\
\normalsize{ Manchester, M13 9PL, United Kingdom }
\\
\normalsize{\up{{7}} European Southern Observatory, Alonso de C\'ordova 3107}\\
\normalsize{ Casilla 19001, Santiago, Chile}
\\
\normalsize{\up{{8}} Isaac Newton Group of Telescopes, Aptdo. de correos 321}\\
\normalsize{ E--38700, S/C de La Palma, Spain}\\
\\
\normalsize{\up{{$^\ast$}} E-mail:  jcorral@iac.es}
}

\date{\footnotesize This is the author's version of the work. It is posted here by permission of the AAAS for personal use, not for redistribution. The definitive version was published in Science on Vol. 339 no. 6123 on 1 March 2013, DOI:10.1126/science.1228222}

\begin{document} 

\hyphenation{si-mi-lar a-zi-mu-thal-ly pro-xi-mi-ty ge-ne-ra-tes o-pe-ra-ted}


\baselineskip24pt


\maketitle 

\newpage


\begin{sciabstract}
Stellar-mass black holes (BHs) are mostly found in X-ray transients, a subclass of X-ray binaries that exhibit violent 
outbursts.  
None of the 50 Galactic BHs known show eclipses, which is 
surprising for a random distribution of inclinations. 
\j13 is a very faint X-ray 
transient detected in 2011.  
On the basis of spectroscopic 
evidence, we show that it contains a BH in a 2.8\,h orbital period. Further, 
high-time resolution 
optical light curves display profound dips without X-ray counterparts. 
The observed properties are best explained by the 
presence of an obscuring toroidal structure moving outwards in the inner disk seen at very high inclination. 
This observational feature should play a key role in models of inner accretion flows and jet collimation mechanisms in stellar-mass BHs.
\end{sciabstract}


Stellar-mass black holes (BHs) are fundamental to our knowledge of supernova explosions and 
stellar evolution \cite{Fryer2001}. 
In $\sim45$ years of X-ray astronomy only 18 BHs have been dynamically 
confirmed in the Milky Way, while $\sim32$ other X-ray binaries 
are suspected to contain a BH due to similar X-ray properties \cite{Ozel2010}.
BHs orbiting massive donor stars can produce X-ray 
eclipses even at moderately low inclinations, 
as exemplified by the bright, persistent extragalactic X-ray binary M33 X-7 
\cite{Orosz2007}. Conversely, none of the $\sim50$ Galactic BH 
transients with low mass donors show eclipses, even though $\sim10$ 
are expected for an isotropic distribution of inclinations \cite{Narayan2005a}.
This implies that selection effects prevent high inclination BH transients from 
being detected, and the most commonly used explanation calls 
for obscuration of the X-rays by a flared accretion disk. Consequently, a large population 
of faint, high-inclination BHs seems to be missed by current X-ray surveys.

The very faint X-ray transient \j13 was detected by the {\it Swift Burst Alert
Telescope} (BAT) in Jan 2011. We used the 2.5\,m Isaac Newton Telescope (INT) 
and the 4.2\,m William Herschel Telescope (WHT) to obtain spectroscopic 
observations of \j13 during the outburst episode between 25 February and 
13 April 2011. The averaged spectra (Fig.~1) 
show a remarkably broad double-peaked \ha~emission line with a full-width at half 
maximum (FWHM) of $\sim3300$\kms. 
The current widest \ha~emission profile is observed in XTE\,J1118+480 
(FWHM $\sim2500$\kms) \cite{Torres2004}, 
an 8\msun~black hole in a 4\,h orbit seen at 
68\degr~inclination \cite{Gelino2006}. 
Larger disk velocities require a 
more massive black hole, a shorter orbital period, a higher inclination 
angle or a combination of these three parameters. 
A double-peaked profile is the signature of gas orbiting in a Keplerian accretion 
disk geometry \cite{Smak1969}, with the peak-to-peak separation driven by the projected velocities 
at the outer edge of the disk. 
From the \ha~profiles we measured a weighted 
average double-peak separation of $1790\pm67$\kms. This can be used to estimate 
the radial velocity semi-amplitude, $K_{\rm c}$, of the companion star in X-ray 
transients \cite{Orosz1994,Orosz1995}, 
yielding a value of $K_{\rm c}\geq 690$\kms~\cite{SM}. Furthermore, the radial velocities 
of the \ha~wings are modulated with a 
$2.8\pm0.3$\,h period (Fig.~1), 
due to the motion of the inner 
accretion disk around the binary's centre of mass. 
The 2.8\,h orbital period,  
combined with our limit to the projected velocity amplitude of the donor star, imply a mass function 
(i.e., an absolute lower limit to the mass of the compact object, \cite{SM})
of $f(M_{\rm x})>3.0$\msun~at a 95.4\% confidence level, and hence, 
the confirmation of a black hole because it exceeds 
the maximum mass possible for a neutron star \cite{Kalogera1996}.

Optical imaging observations were taken using the 0.82\,m IAC80, the 1.2\,m Mercator (MT), the 
2.0\,m Liverpool (LT) and the INT telescopes on 16 nights between March and July 
2011 (Table S1) 
during the smooth decay from outburst.
We used very short exposure times on six of the nights, resulting in time resolutions ranging from 7--22\,s. This was crucial 
to reveal the fast variability structure in the light curves, 
which unveils the presence of striking dipping variability (Fig.~2). 
Further, the high-time resolution allowed us to resolve the complex structure of the dip profiles into 
multiple irregular eclipse-like features (see Fig.~2c). 
In order to explore the timing 
properties of the dips, we computed Lomb-Scargle periodograms of the high-time resolution data (Fig.~3). 
The Power Density Spectra (PDS) reveal a strong peak produced by the dip recurrence 
period (DRP) which migrates from 2.3 to 7.5\,min over our 69\,d baseline. 
We found no evidence for modulation on the 2.8\,h~orbital period over the entire photometric database.

The dipping activity in \j13 is reminiscent of neutron star X-ray dippers 
\cite{White1985}, where the dips are  
interpreted as the occultation of the central X-ray source by an azimuthally structured outer 
disk rim seen at $\sim70$\degr~inclination \cite{White1985}. However, 
unlike the behaviour of known X-ray dippers, the short repeatability 
time scale observed in \j13 ($\sim2-8$\,min,  i.e. a tiny fraction of the orbital period) 
and the changing recurrence period of the optical dips indicate that the eclipsing material 
is not phase locked to the disk rim. Instead, the dips in \j13 can only be produced by fast 
and regular occultations of bright optically thick regions by a vertically extended structure 
well within the disk. Further, the (FWHM) duration of the dips is $\sim50$\% of their recurrence
time, implying that the obscuring geometry extends $\sim180$\degr, i.e. an inner disk torus 
of variable height (perhaps tilted with respect to the binary plane) or a warp. 

Several scenarios can in principle explain the formation of inner disk structures. 
Numerical simulations performed for extreme mass ratio systems show that 
X-ray irradiation by the central source can generate a torque that triggers a twist or warp \cite{Foulkes2006}.
The precession of the warp generates a superhump modulation at approximately the orbital 
period, i.e. substantially longer than our DRP. Alternatively, the Bardeen-Petterson effect 
around a rapidly rotating compact object can cause a tilted accretion disk to warp into the 
equatorial plane at a few tens $R_g$ and naturally produce photometric modulations \cite{Bardeen1975}. 
However, under the assumption that the DRP reflects the Keplerian frequency of the obscuring wall,
this would be placed at much larger distances ($\sim10^3 - 10^4\,R_g$) than predicted by Lense-Thirring 
simulations \cite{Lense1918,Ingram2009}. Also, our observations of an increasing DRP
provides the signature of a travelling wave propagating outwards  
(Fig.~3).

The detection of optical dips requires the observer's line-of-sight to be close to the 
plane of the binary orbit; the remarkable dipping amplitude of up to 
0.8\,mag implies grazing eclipses of a compact source with a 50\% reduction in flux,  
 which suggest large inclinations $i\gtrsim 70$\degr. 
Why doesn't the companion star eclipse the optical source at these high inclinations? 
Unlike neutron star X-ray binaries, \j13 
has a very extreme mass ratio $q=M_{\rm c}/M_{\rm x}<0.06$, 
where $M_{\rm c}$ and $M_{\rm x}$ stand
for the mass of the companion star and the compact object, respectively \cite{SM}.
Assuming a typical disk extending up to the tidal truncation radius \cite{King1995} and 
with scale height $\alpha=12$\degr~\cite{deJong1996}, the small $q$ results in a donor star 
radius comparable to or smaller than the disk's outer rim.
Thus, the central disk regions are never eclipsed by the donor star even at the 
extreme scenario of an edge-on geometry (Fig.~S3).
In addition, regardless of the inner disk's elevation, the donor star is 
always sheltered from irradiation by the outer disk, preventing the production of orbitally 
modulated optical light curves.
We estimate that the donor is consistent with an slightly
evolved $\textrm{M4.5}$ star with $M_{\rm c}=0.24$\msun~and  
$R_{\rm c}=0.29$\rsun~\cite{SM}.

To further constrain the binary inclination we searched for the presence of X-ray 
eclipses/dips in the public RXTE archive between 2 February and 2 April 2011, but could not 
find any \cite{SM}. 
Eclipses of X-ray photons scattered by a disk corona are typically seen 
in neutron star X-ray binaries at high inclination $i\gtrsim 80$\degr 
\cite{Frank1987,vanParadijs1988}. However, these may be
too shallow and hardly detectable in \j13, given the very small size of the  
donor star and the inner disk torus compared to that of a putative disk corona.   
We also note that no signature of dips, eclipses or
modulations was detected in \textit{Swift}/XRT data \cite{Armas-Padilla2012}.

At a distance of $\sim1.6$\,kpc \cite{SM}, \j13 has an unusually weak X-ray peak luminosity 
$L_{\rm x}\sim1.7\times10^{35}$ erg s$^{-1}$. The luminosity ratio 
$L_{\rm x} (2-11\,{\rm keV})/ L_{\rm opt} (3000-7000\,{\rm \AA})\sim 57$ is also 
atypically low for an X-ray transient in outburst \cite{vanParadijs1995}. 
These properties are reminiscent of Accretion Disk Corona sources, 
where the central engine is hidden from view and only X-rays scattered into our line of sight 
by material above the disk plane can be seen. 
The unique features of \j13 (namely, its very low X-ray luminosity, striking optical dips, 
lack of orbital modulation  and extremely broad \ha~profile) can thus be explained by orientation 
effects in an edge-on geometry. 
However, at variance with the commonly accepted scenario, based on Milgrom's model 
\cite{Milgrom1978}, our observations reveal that the central black hole is hidden by an inner 
toroidal structure rather than the outer disk rim. 
Orientation effects are also invoked to explain the observable properties of 
active galactic nuclei (AGN), where Seyfert 2 objects are thought to be edge-on 
AGNs obscured by an opaque torus \cite{Antonucci1993}. In this 
framework, it is tempting to view \j13 as a scaled-down version of the Seyfert 
2 scenario, where distinct observable   
properties are a consequence of geometrically thick inner disk structures seen at very high inclination.

Inner disk donuts might be ubiquitous in erupting X-ray binaries, at least during the hard 
state when there is emission from compact jets (see \cite{McClintock2006a} for a review on BH states) - 
a state where \j13 remained through the entire outburst \cite{SM,Armas-Padilla2012,Krimm2011b,Sivakoff2011}. 
This observational feature may be a key to understand 
the poorly known accretion-ejection connection and the collimation of jet 
outflows \cite{Livio1999} observed during the transition from the hard to the 
soft state. Standard mechanisms for jet production require both high angular velocities and strong 
poloidal magnetic fields which, in turn, depend on the disk height. 
Geometrically thick, inner disk regions such as that observed in \j13 
provide the best environment for jet formation \cite{Meier2001}. 
Note that a radio detection of \j13 was reported during outburst, presumably associated 
to compact jet emission \cite{Sivakoff2011}.
In this regard, it has been proposed that (irradiation induced) equatorial disk winds 
develop during BH soft states but not in the jet/hard X-ray state \cite{Ponti2012}. 
The presence of inner vertical structures is relevant to this picture   
because it prevents a large irradiation of the outer disk, thus providing an explanation for the absence of equatorial winds in the hard state. This scenario 
is consistent with the results reported in \cite{Armas-Padilla2012}, which imply that the accretion 
disk in \j13 is not irradiated.

We note that the detection of an edge-on black hole has been possible due to its proximity.
With a maximum flux of 30\,mCrab in the 15--50\,keV band, \j13 
was barely detected by \textit{BAT} on-board \textit{Swift} \cite{Krimm2011a}; were it placed just farther than $\sim1.9$\,kpc, it would have not reached a $3\sigma$ detection. In addition, with a Galactic latitude $b=+50$\degr~it is 
located in the halo at $\sim1.2$\,kpc above the Galactic plane, where the 
absorption column $N_{\rm H}$ is minimal. Other edge-on black holes sitting on 
the Galactic plane would certainly be affected by interstellar extinction, which 
makes their outbursts difficult to detect by soft X-ray wide field cameras 
like ASM/RXTE.   

Some of these may be associated to the increasing population of Very Faint X-ray Transients 
recently detected \cite{King2006}. Clearly, the fortuitous short distance and 
the dramatic optical dips have led to the detection and characterization of \j13, a prototype of the hitherto 
missing population of high-inclination BH X-ray transients. 


\nocite{Paczynski1971}\nocite{Naylor1998}\nocite{Calvelo2009}\nocite{Casares1995b}
\nocite{Harlaftis1999}\nocite{dellaValle1993}\nocite{Filippenko1999}\nocite{Smak1984}
\nocite{Baptista2001}\nocite{Schneider1980}\nocite{Casares2011}\nocite{Shafter1986}
\nocite{Rau2011}\nocite{Warner1995}\nocite{Smith1998}\nocite{Smith1998}\nocite{Knigge2006}
\nocite{Knigge2011}\nocite{Motta2011}

\bibliography{1228222Revisedtext}

\bibliographystyle{Science}

\noindent


\section*{Acknowledgement}
 This work has made use of the \textsc{iraf} facilities and the \textsc{molly} software developed 
by T.~R.~Marsh. 
 We want to thank P. Charles, G. Ponti and D. M. Russell for their useful comments,
G. P\'erez and M. Leal for their help with the artistic figure and I. Negueruela and 
C. Gonz\'alez-Fern\'andez for obtaining the IDS spectra.
 Based on observations made with: the INT and WHT operated by the Isaac Newton Group, 
the Liverpool Telescope operated by the Liverpool John Moores University with financial support from 
the UK Science and Technology Facilities Council, and the Mercator telescope 
operated by the Univ. of Leuven and the Obs. of Geneva. All of them 
installed at the Spanish Observatorio del Roque de Los Muchachos 
(on the island of La Palma) of the Instituto de Astrof\'isica de Canarias (IAC).
 Based on observations made with the INT Telescope under the Spanish 
IAC Director's Discretionary Time.
TMD acknowledges funding via an EU Marie Curie Intra-European Fellowship under 
contract no. 2011-301355.
This research has been supported by the Spanish Ministerio de Ciencia e Innovaci\'on 
(MICINN) under grants AYA2010--18080. Partly funded by the Spanish MICINN under the 
Consolider-Ingenio 2010 Program 
grant CSD2006--00070: ''First science with the GTC'' 
(\url{http://www.iac.es/consolider-ingenio-gtc/}).\\

\noindent
{\small
\textbf{Supplementary Materials}\\
www.sciencemag.org\\
Supplementary Text\\
Figs. S1, S2, S3\\
Table S1\\
References (33-50)\\
Movie S1}


\clearpage


\pagestyle{empty}

\begin{figure}
    \includegraphics[width=\textwidth]{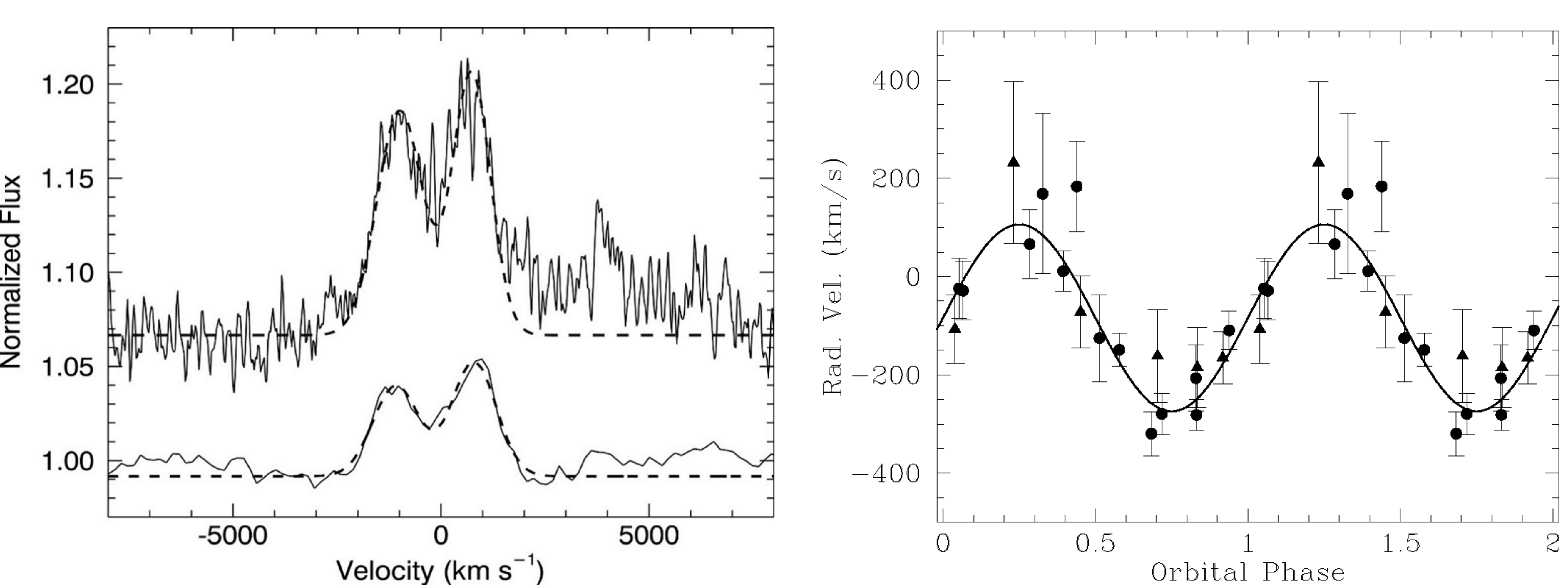}
    \caption{
    \textbf{
    Average \ha~emission line profiles and radial velocity curve of the 
    \habf~emission line folded on the 2.8\,h period. Left:} The line weakens from $\textrm{EW}=6-9$\,\AA~on 
    25--27 February to 2.1\,\AA~on 19 March 2011. A vertical offset has been applied for clarity. The 
    \ha~line is the only strong emission feature detected in the optical
    spectra. 
    Two Gaussian fits to the double-peaked profiles are
    also shown (dashed lines).
    \textbf{Right:} The velocities were obtained using a double Gaussian passband with a 
    separation of 2600\kms. Phase zero corresponds to HJD\,2455640.588. Solid circles indicate 
    ACAM velocities from 19 March and solid triangles IDS velocities on 25--27 February. 
    The best sine fit is shown (solid line).
    }
\end{figure}
\newpage
\thispagestyle{empty}
\begin{figure}[t]
    \vspace*{-1.0cm}
    \includegraphics[width=\textwidth]{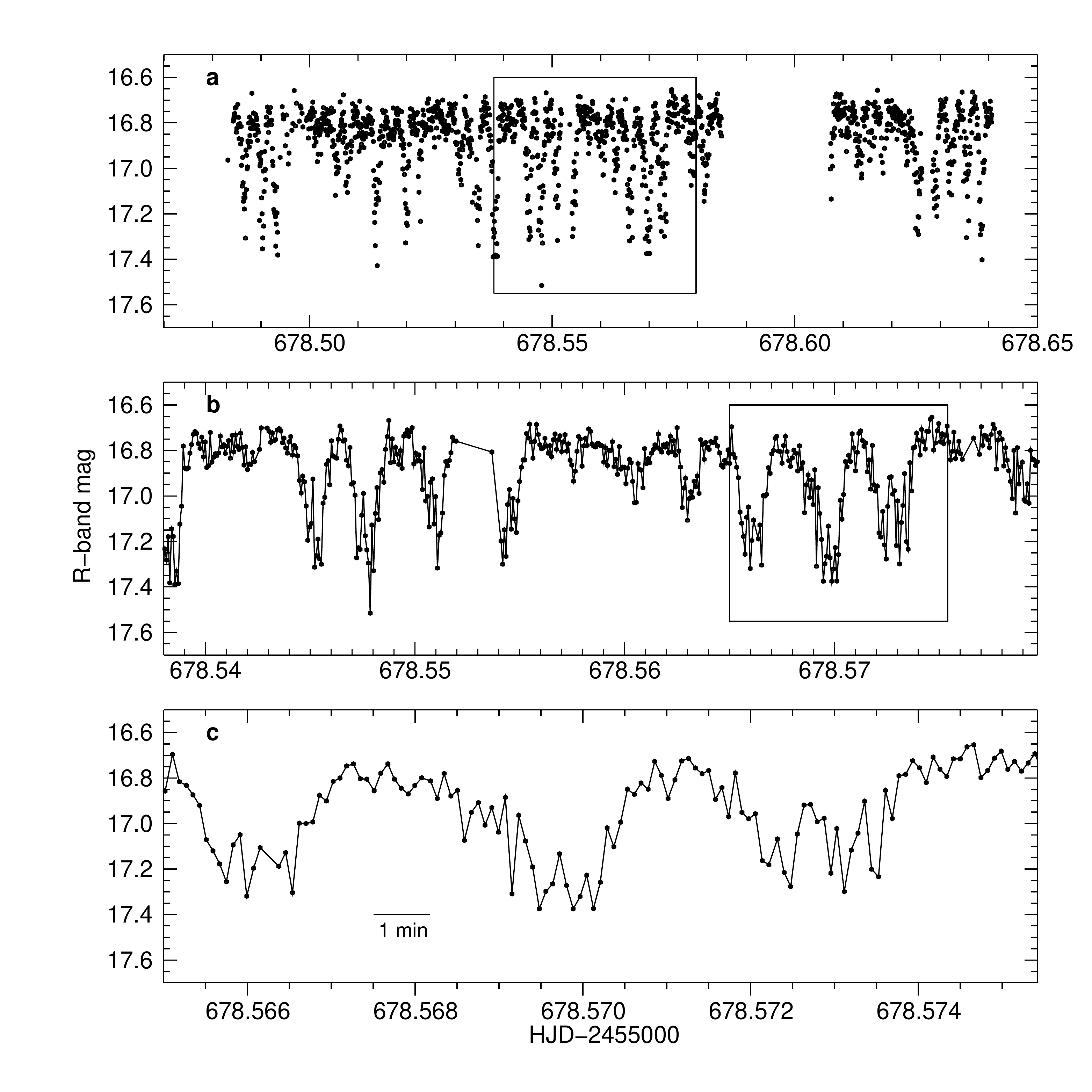}
    \caption{\small
  \textbf{
  The optical light curve of \j13 taken on 26 April 2011 with the INT}. 
  \textbf{a}: The time resolution is 7\,s whereas the total duration of the observing run is 3.8\,h. 
  Regular dips cause a drop in brightness of up to $\simeq0.8$\,mag.
  \textbf{b}: A close up of the box indicated in \textbf{a} with a length of 1\,h. Dips have a 
  characteristic (FWHM) size of 2\,min and recur every 4\,min, hence their duty cycle is $\sim50$\%, 
  substantially larger than the typical 10--30\% duty cycles observed in neutron star dippers. 
  The depth of the dips also appears modulated with a time scale of $\sim30$\,min. 
  \textbf{c}: A 15\,min close up of the box indicated in \textbf{b}.  
  The complex structure of the dip profiles indicates that the obscuring region is 
  clumpy and has considerable structure. Large intensity drops of $0.4$\,mag ($\sim30$\% 
  reduction in flux) are seen on time-scales not resolved by our 7\,s time resolution implying 
  that the eclipsed region is very compact. Alternatively, it may be composed of a distribution 
  of small emission knots. Assuming that the DRP is driven by the orbital frequency 
  of a vertical disk annulus around a 10\msun~black hole, its Keplerian velocity is 
  $\sim3150$\kms. Our 7\,s time resolution, therefore, sets an upper limit to the size of the 
  eclipsed optical region (or individual emitting knots) of $\sim0.03$\rsun. This is $\sim2$\%  
  the size of a typical accretion disk extending to the tidal truncation radius \cite{King1995} 
  in a 2.8\,h black hole binary. Each of our 16 night light curves displays 
  similar dipping activity, but it is only resolved in detail on the 6 nights sampled at 
  high-time resolution.
  }
\end{figure}
\thispagestyle{empty}
\newpage
\thispagestyle{empty}
\begin{figure}
    \includegraphics[width=\textwidth]{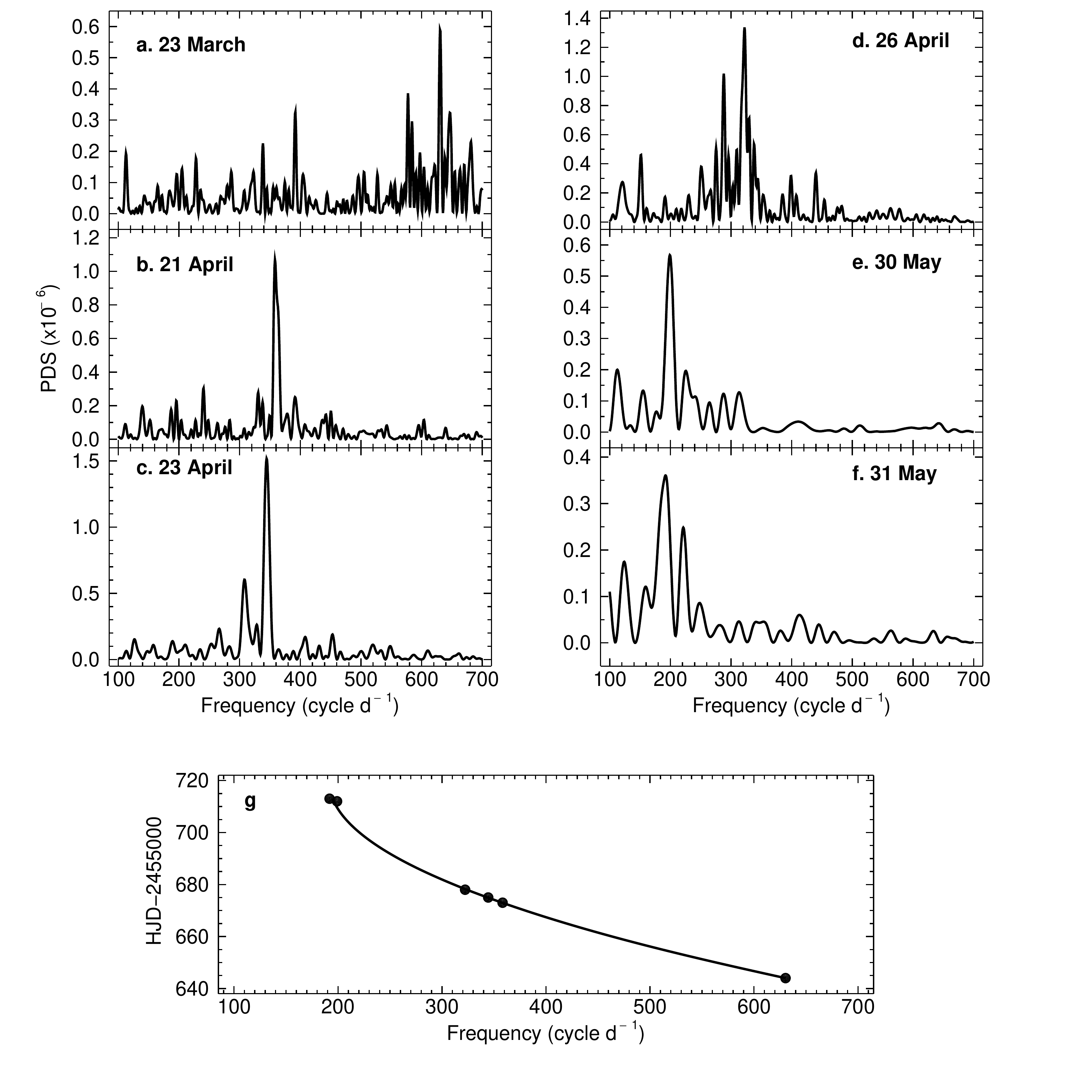}
    \caption{
   \textbf{
   Time evolution of the optical dip recurrence period (DRP).} 
   Panels \textbf{a}--\textbf{f} display 
   the power density spectra (PDS) of the 6 high-time resolution nights. The highest 
   peak shifts from $2.3$\,min ($\sim630$\,cycle\,d$^{-1}$) to $7.5$\,min ($\sim192$\,cycle\,d$^{-1}$) 
   over 69\,d. Panel \textbf{g} shows the frequency shift as a function of time with the best parabolic 
   fit (solid line): $\nu_{\rm DRP}=7.7\times 10^{-2}\,T^2 - 111.218\,T + 40192.4$, with $\nu_{\rm DRP}$ 
   the DRP frequency, $T=\textrm{HJD}-2455000$ and HJD representing the Heliocentric Julian Date of the 
   observation. Assuming that the DRP period reflects the Keplerian frequency of a particular disk annulus 
   in the gravitational field of a 10\msun~black hole, this frequency corresponds to just $\simeq$1600 
   Schwarzschild radii. Further, the 2.3\,min DRP on 23 March is produced at 0.12\rsun, while the $7.5$\,min 
   DRP on 31 May is at 0.27\rsun. Consequently, the disk annulus responsible for the DRP has moved outwards 
   by 0.15\rsun~in $69$\,d at an average speed of $17$\,m~s$^{-1}$. 
   }
\end{figure}

\clearpage



\section*{Supplementary material}

# Supplementary Text

This Supplementary Material provides details on the acquisition and reduction of the spectroscopic and photometric data, including a complete log of all the observations. It also contains details on some information used in the main body of the paper, in particular the determination of the orbital period and constraints to the mass function and binary mass ratio. Also, the distance to the binary is derived based on a semi-empirical donor sequence for close binary stars. Finally, we present the analysis of RXTE light curves obtained nearly simultaneously to our optical observations.

## Spectroscopic observations

Six 30–33 min spectra of *Swift* J1357.2−0933 were obtained on the nights of 25–27 February 2011 using the Intermediate Dispersion Spectrograph (IDS) on the Isaac Newton Telescope (INT) at the Observatorio del Roque de los Muchachos on La Palma (Spain). The observations were performed with the H1800V grating and a slit width of 1.6 arcsec to yield a spectral coverage of 6270 – 7000 Å with 30 km s$^{-1}$ FWHM resolution at H$\alpha$. The IDS spectra were reduced using standard techniques including debiasing and flat field correction. The spectra were subsequently extracted using optimal extraction techniques within the PAMELA package. A spectrum of a CuNe arc lamp was obtained after every target exposure and the pixel-to-wavelength calibration was derived from a 3rd order polynomial fit to 11 arclines which resulted in a rms scatter of 0.014 Å.

We also used the Auxiliary-port CAMera (ACAM) on the 4.2 m William Herschel Telescope (WHT) to obtain lower resolution spectroscopy of *Swift* J1357.2−0933 during service time. ACAM is mounted permanently at the folded Cassegrain focus of the WHT with the V400 VPH grating, which provides spectroscopy with resolving power $R \sim 450$ in the range $\lambda\lambda 4400-9200$. We obtained 61 and 14 spectra on the nights of 19 March and 13 April 2011, respectively. A 1 arcsec slit was used on both nights which resulted in ∼ 560 km s$^{-1}$ resolution at H$\alpha$, as measured from the FWHM of the arc lines. The integration time was set to 250 s on the first run but it was increased to 600 s on the second due to seeing degradation. Standard procedures were followed within IRAF to de-bias and flat-field correct the ACAM spectra. In addition to the poorer weather conditions we find that the H$\alpha$ emission weakens dramatically to an equivalent width (EW) ∼ 0.1Å on the second ACAM night. Hence, only spectra from IDS and the first ACAM night are used in the analysis. Wavelength calibration was obtained by fitting 24 CuNe arc lines with a 3rd order Legendre function. The fit resulted in a dispersion of 3.4 Å pix$^{-1}$ and an rms scatter of 0.13 Å. Several arcs were obtained along each night and the fitted coefficients were interpolated to the times of the target observations. The instrumental flexure was also monitored by measuring the position of the sky line OI 5577.34 Å and the resulting offsets were removed from the individual spectra.

**Photometric observations**

*R*-band images of *Swift* J1357.2−0933 were collected between 23 March and 31 July 2011 using the Merope CCD camera on the 1.2 m Mercator Telescope (MT), the RATCam camera on the 2.0 m Liverpool telescope (LT) and the Wide Field Camera (WFC) on the INT. All these telescopes are operated at the Roque de los Muchachos Observatory. Further observations were also obtained with the CAMELOT camera attached to the 0.82 m IAC80 telescope at the Observatorio del Teide between 2 March and 18 June 2011. The log of the photometric observations is presented at the top of Table S1. Integration times between 3 and 60 s were employed depending on telescope size and weather conditions. A small window was applied to the WFC detector CCD#4 on 21–26 April, which resulted in a time resolution of 7–8 s. All images were taken through Harris or Bessell *R*-band filters except for the last data point on 31 July 2011 when the Sloan-*r'* filter was used. Images were corrected for bias and flat-fielded in the standard way using IRAF. We applied optimal aperture photometry (34) to *Swift* J1357.2−0933 and 5 comparison stars close to the target. These stars were calibrated during the first IAC80 campaign under photometric conditions, using the photometric transformation developed by the IAC80 group (see http://www.iac.es/telescopes/pages/en/home/utilities.php for further information).

**Analysis of the spectroscopy**

The extreme width of the Hα emission line (Fig. 1), with FWHM ∼ 3300 km s$^{-1}$, is remarkable. A double Gaussian fit to the emission profile yields a double peak separation of 1726 ± 83 km s$^{-1}$ and 1911 ± 114 km s$^{-1}$ for the IDS and ACAM averaged spectra, respectively.

The separation of the peaks in the Hα profile can be used to estimate the projected velocity $K_c$ of the companion star in X-ray transients. It has been shown (*8, 9*) that $v_d/K_c \simeq 1.1 - 1.25$ for quiescent black hole transients, where $v_d$ stands for the outer disk velocity or half the double peak separation. The weighted mean of the double peak separation in the IDS and ACAM spectra is 1790 ± 67 km s$^{-1}$. Therefore, $v_d = 895 \pm 33$ km s$^{-1}$, and hence, a conservative lower limit of $K_c = 716 \pm 26$ km s$^{-1}$ (or $K_c > 690$ km s$^{-1}$) can be derived from the most extreme value $v_d/K_c = 1.25$. One should note that $K_c$ is very likely underestimated because the empirical relation was obtained from quiescent accretion disks (*9*), whereas *Swift* J1357.2−0933 is in outburst. The double peak velocity separation is known to be smaller during outburst than in quiescence because the disk expands by the action of the viscous instability. For example, the double peak separation in XTE J1118+480 is ∼ 1200 km s$^{-1}$ in outburst and ∼ 1700 km s$^{-1}$ in quiescence, that is, a factor ∼ 1.4 larger (*35*). Other examples are provided by GRO J0422+320, with $v_d \sim$ 800 km s$^{-1}$ in outburst (*36*) and ∼ 1000 km s$^{-1}$ in quiescence (*37*) or Nova Vel 93, with ∼ 780 km s$^{-1}$ in outburst (*38*) and ∼ 1200 km s$^{-1}$ in quiescence (*39*). Further evidence for the outburst expansion of accretion disks is provided by the changing width of the hot spot eclipse in U Gem (*40*) and eclipse mapping of dwarf novae (*41*). In summary, we decided to take 690 km s$^{-1}$ as a very conservative lower limit to the velocity semi-

amplitude of the donor star in *Swift* J1357.2−0933.


**Orbital period**

We have searched for evidence of the orbital period in *Swift* J1357.2−0933 by inspecting the radial velocities of the Hα emission line. The 61 spectra of the first ACAM night were co-added in groups of four in order to increase the signal-to-noise ratio, but two were rejected because of poor statistics. These, together with the 6 IDS spectra, leave a total of 19 spectra for the analysis.

To avoid contamination caused by variability in the double peak, radial velocities were extracted from the line wings using the double-Gaussian technique (*42*). Every Hα profile was convolved with a template consisting of two Gaussian bandpasses with FWHM 300 km s$^{-1}$ and separations in the range $a = 2000 - 3000$ km s$^{-1}$. After several tries we decided to take $a = 2600$ km s$^{-1}$ as the optimal value that provides the cleanest radial velocity curve. A ∼ 3 h modulation is already apparent in the radial velocities of the ACAM night. A $\chi^2$ periodogram analysis of all the radial velocity points displays a clear minimum at 8.576 cycle d$^{-1}$ or 0.1166 d (see Fig. S1). A conservative estimate of the period uncertainty is provided by the 99% confidence level in the periodogram. This yields 2.8 ± 0.3 h as our best estimate of the orbital period of *Swift* J1357.2−0933. We note that this is close to the ∼ 2 h period previously reported under the assumption that the donor is a main sequence M4 star filling its Roche lobe (*43*).


**Constraints on stellar masses and binary distance**

The mass function $f(M_x) = P_{orb} K_c^3 / 2 \pi G = M_x \sin^3 i / (1 + q)^2$ provides an absolute lower limit to the mass of the compact object $M_x$, which is obtained when the companion's mass $M_c$ is neglected and the binary is seen edge-on ($i = 90°$). The radial velocity analysis yields an orbital period of 2.8 ± 0.3 h, whereas the double peak separation implies $K_c = 716 \pm 26$ km s$^{-1}$. Propagating errors in the mass function equation yields $f(M_x) > 3.0$ M$_\odot$ with a 95.4% confidence, and hence, strongly advocate for the presence of a black hole in *Swift* J1357.2−0933. As we showed in the previous section, $K_c$ is likely underestimated which even reinforces the case for a black hole.

To constrain the orbital motion of the compact star we extracted radial velocities from the Hα emission line using the double-Gaussian technique for a range of Gaussian separations between $a = 2100 - 3900$ km s$^{-1}$ in steps of 100 km s$^{-1}$. Only spectra corresponding to the ACAM night of 19 March were considered, which were co-added into 15 phase bins using the 2.8 h orbital period derived above. The zero phase was arbitrarily set to HJD 2455640.588, as given by the sine wave fit for $a = 2600$ km s$^{-1}$. Each radial velocity curve was fitted with the expression $V(\phi) = \gamma + K_1 \sin 2\pi (\phi - \phi_0)$, and the evolution of the fitting parameters $K_1$, $\gamma$ and $\phi_0$ versus the Gaussian separation $a$ is displayed in Fig. S2 as a diagnostic diagram. The high velocity wings are formed in the inner regions of the accretion disk and hence their radial velocity curve is expected to

3

trace the motion of the compact star. As we move away from the line core, the radial velocity semi-amplitude $K_1$ decreases from $\sim 250$ km s$^{-1}$ to $\sim 48$ km s$^{-1}$ at $a = 3600$ km s$^{-1}$. Beyond $a = 3600$ km s$^{-1}$ $K_1$ remains stable in the range 38 – 48 km s$^{-1}$, with a mean value of $43 \pm 2$ km s$^{-1}$. This is also when the diagnostic parameter $\sigma(K_1)/K_1$ starts to rise, which marks the limit when the continuum noise begins to dominate (*44*). Therefore, we decide to take $K_1 = 43 \pm 2$ km s$^{-1}$ as the best description of the radial velocity semi-amplitude of the compact star. Combined with our lower limit to $K_c$ this results in a binary mass ratio $q = K_1/K_2 < 0.06$ which is typical of black hole X-ray binaries.

It has been reported that the $(i - z)$ color and apparent magnitude of the quiescent counterpart to *Swift* J1357.2−0933 are consistent with an unevolved M4 star at a distance of $\sim 1.5$ kpc (*45*). If the donor star were a M4 main sequence, the period-density relation for a Roche lobe filling star (*46*) would result in a $\sim 2$ h orbital period (*43*). Our detection of a 2.8 h orbital period in turn would imply a M2 main sequence star. However, donor stars in short period interacting binaries deviate from main sequence stars because they are not in thermal equilibrium. Consequently, we use empirical relations (*47*) to describe the physical structure of the companion star in *Swift* J1357.2−0933. For an orbital period of 2.8 h these relations yield $M_c = 0.24 \pm 0.07$ M$_\odot$, $R_c = 0.29 \pm 0.03$ R$_\odot$ and a spectral type M4.5 $\pm$ 0.8, in good agreement with the quiescent colors (*45*). As expected by its departure from thermal equilibrium, the donor in *Swift* J1357.2−0933 has a later spectral type than a main sequence star in a 2.8 h period (*48*). The estimated donor's mass, combined with our previous lower limit $M_x > 3.0$ M$_\odot$ supports again an extreme mass ratio $q = M_2/M_1 < 0.06$.

We now use the semi-empirical donor sequence developed for cataclysmic variables (*49*) to estimate the absolute magnitude of a M4 star with $R_c \simeq 0.29$ R$_\odot$ and get $M_R \simeq 10.8$. Assuming that the quiescent light is dominated by the donor star, the distance modulus equation and the apparent magnitude $r = 21.96$ (*45*) imply a distance $d \sim 1.6$ kpc, where an interstellar extinction $E(B - V) = 0.04$ has been adopted (*45*). Note that, strictly speaking, this is a lower limit to the distance because we have neglected any contribution from the accretion disk light to the quiescent magnitude, although the observed colors (*45*) do suggest this is a fair approximation.

## **X-ray data analysis**

We analysed 25 observations of *Swift* J1357.2−0933 taken with the *Proportional Counter Array* (PCA) on-board *Rossi X-ray timing explorer* (RXTE). Observations were performed between 2 February and 2 April 2011. For each observation we extracted background and dead-time corrected energy spectra and light curves using the standard RXTE software within HEASOFT V. 6.12. We computed light-curves within the energy band $\sim 2 - 15$ keV using 1 s time resolution but could not detect the presence of dip-like or eclipse-like structures in the data. This is also consistent with higher sensitivity studies performed by *Swift*/XRT (*21*) and XMM.

The PCA Standard 2 mode (STD2) was used for spectral analysis. We extracted the

spectral hardness ($h$), defined as the ratio of counts in STD2 channels 11 – 20 (6.1–10.2 keV) and 4 – 10 (3.3–6.1 keV). Even though only half of the observations show a count rate high enough to perform a proper hardness study ($\geq 15$ cts s$^{-1}$), the values obtained ($h \sim 0.80 - 0.85$) are consistent with those typically observed during hard states in black holes candidates (*50*).

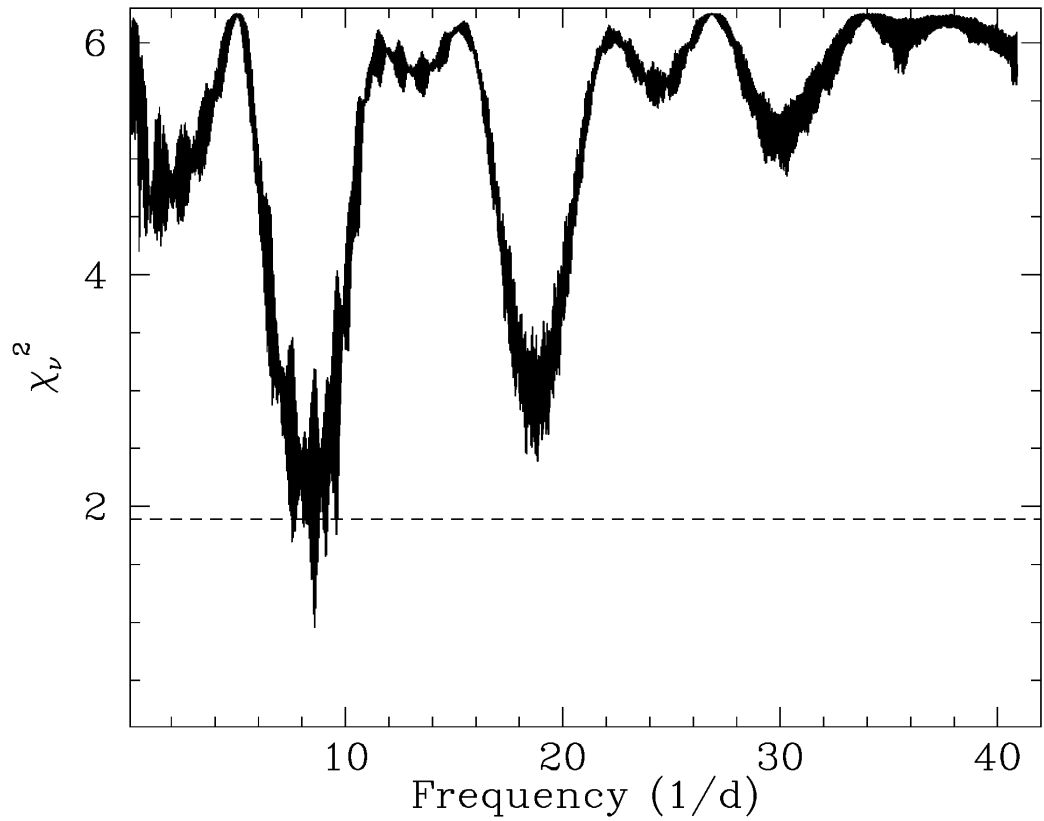

**Fig. S1. $\chi^2$ periodogram of the Hα radial velocities from IDS and the ACAM night on March 19.**

The deepest peak corresponds to 8.576 $d^{-1}$. The dashed line marks the 99% confidence level.

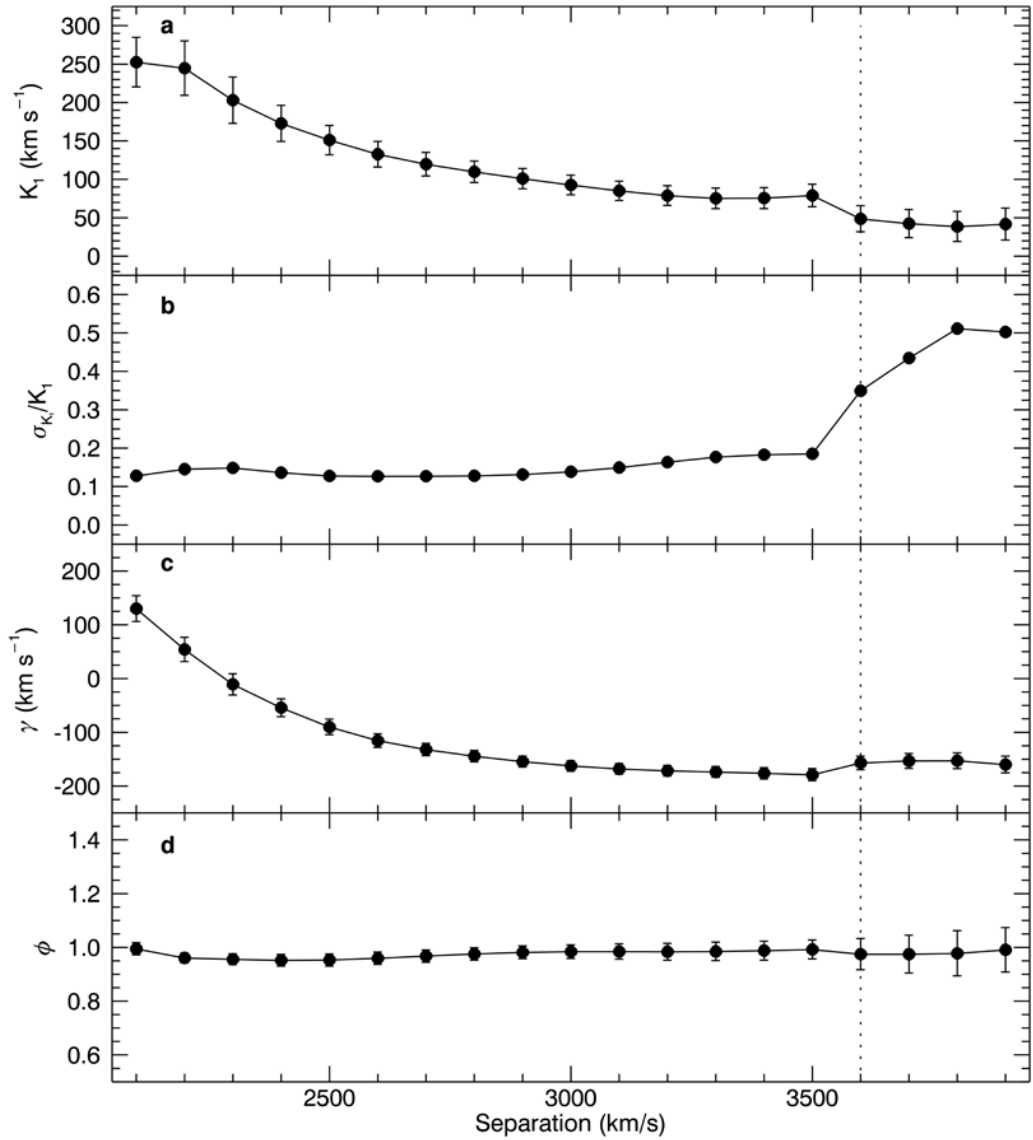

**Fig. S2. Diagnostic diagram for Hα.**
It is calculated using the double Gaussian technique with FWHM=350 km s$^{-1}$. The vertical line marks the Gaussian separation where the continuum noise starts to dominate the radial velocities.

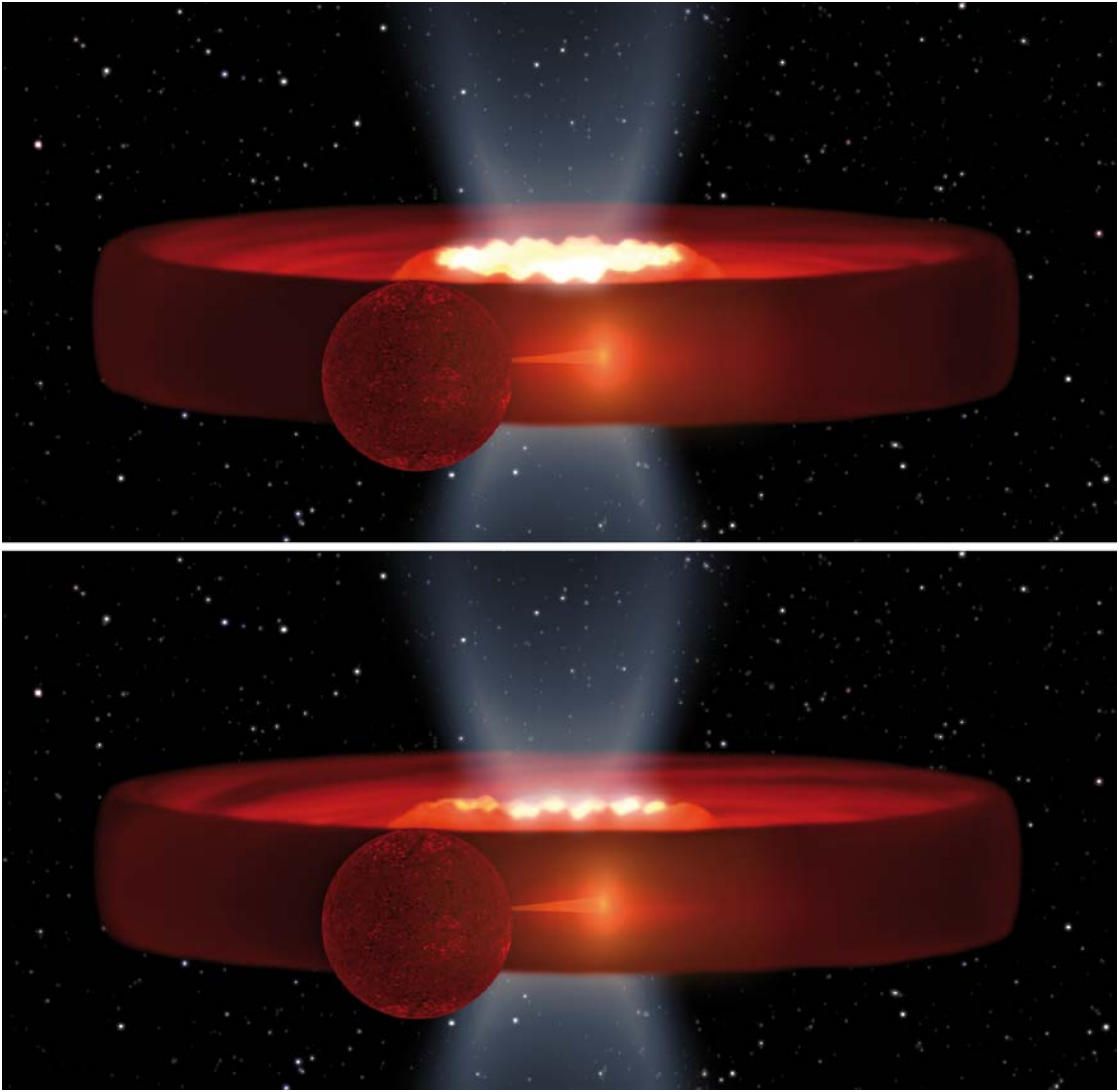

**Fig. S3. An artist's impression of *Swift* J1357.2−0933 seen at 85º inclination.**
Our observations of profound rapid optical dips imply a geometrically thick inner warp/torus eclipsing a very compact or clumpy optically thick region located above the disk plane. The two panels show opposite phases of the dip recurrence period corresponding to the inside (bottom) and outside dip light curve (top). All the binary elements are shown to scale, with a binary mass ratio $q = 0.024$ and a disk of radius 0.9 $R_{L1}$ and 12º flaring angle. $R_{L1}$ is the effective radius of the primary Roche lobe i.e. $R_{L1} = a - R_2$ with $R_2$ the effective radius of the donor star and $a$ the binary separation. The latter is provided by Kepler's third law $a^3 = G/4 \pi^2 (M_x + M_c) P_{orb}^2$, where $M_x$ and $M_c$ are the masses of the black hole and the companion star, respectively. We also used the expression $R_2/a \simeq 0.462(q/(1+q))^{1/3}$ which is accurate to within 2% for small mass ratios $q = M_c/M_x < 0.8$ (33). In the case shown we have adopted $M_x = 10$ M$_\odot$, $M_c = 0.24$ M$_\odot$ and $P_{orb} = 2.8$ h, which results in $a = 2.18$ R$_\odot$, $R_2/a = 0.13$ and $R_{L1} = 1.90$ R$_\odot$. Note that, for this extreme mass ratio and inclination, the donor star only eclipses the disk rim. Since this rim is very cold, its contribution to the optical flux is negligible.

**Table S1. Log of the photometric and spectroscopic observations.**

**Top:** Photometric observations. Δt (column 4) lists the total on-source time, while δt (column 5) is the effective time resolution (including CCD readout). Magnitudes are in the *R*-band except where indicated. **Bottom:** Spectroscopic observations.

| Date (2011) | HJD -2455000 | Telescope+instrument | Δt (h) | δt (s) | # Exp. | Aver. magnitude |
|---|---|---|---|---|---|---|
| Mar. 02 | 623 | IAC80+CAMELOT | 4.15 | 85 | 41 | 16.57 ± 0.04 |
| Mar. 23 | 644 | IAC80+CAMELOT | 0.92 | 85 | 44 | 16.66 ± 0.19 |
| Mar. 23 | 644 | INT+WFC | 4.78 | 17 | 847 | 16.70 ± 0.11 |
| Mar. 25 | 646 | LT+RATCAM | 3.98 | 71 | 199 | 16.62 ± 0.09 |
| Mar. 26 | 647 | LT+RATCAM | 3.98 | 72 | 199 | 16.61 ± 0.08 |
| Mar. 29 | 650 | IAC80+CAMELOT | 3.22 | 85 | 131 | 16.79 ± 0.07 *a* |
| Apr. 09 | 661 | IAC80+CAMELOT | 1.06 | 85 | 41 | 16.82 ± 0.08 |
| Apr. 15 | 667 | IAC80+CAMELOT | 4.12 | 85 | 139 | 16.75 ± 0.11 |
| Apr. 20 | 672 | IAC80+CAMELOT | 2.22 | 85 | 92 | 16.80 ± 0.09 |
| Apr. 21 | 673 | INT+WFC | 4.11 | 8 | 1820 | 16.84 ± 0.12 |
| Apr. 23 | 675 | INT+WFC | 2.46 | 8 | 949 | 16.84 ± 0.14 |
| Apr. 26 | 678 | INT+WFC | 3.77 | 7 | 1483 | 16.87 ± 0.14 |
| May. 05 | 687 | IAC80+CAMELOT | 1.23 | 85 | 51 | 16.89 ± 0.08 |
| May. 30 | 712 | MT+MEROPE | 1.26 | 22 | 206 | 17.34 ± 0.10 |
| May. 31 | 713 | MT+MEROPE | 1.50 | 22 | 214 | 17.32 ± 0.11 |
| Jun. 18 | 731 | IAC80+CAMELOT | 0.25 | 85 | 10 | 17.14 ± 0.03 |
| Jul. 31 | 774 | INT+WFC | — | — | 1 | 17.59 ± 0.17 *b* |

| Date (2011) | HJD -2455000 | Telescope+instrument | Wavel. range (Å) | Dispersion (Å pix$^{-1}$) | Exp.time (s) |
|---|---|---|---|---|---|
| Feb. 25 | 618 | INT+IDS | 6270–7000 | 0.66 | 2x1800 |
| Feb. 26 | 619 | INT+IDS | " | " | 2x2000 |
| Feb. 27 | 620 | INT+IDS | " | " | " |
| Mar. 19 | 640 | WHT+ACAM | 4400–9200 | 3.4 | 61x250 |
| Apr. 13 | 665 | WHT+ACAM | " | " | 14x600 |

*a* Taken with a *V* filter.
*b* Taken with a Sloan-*r* filter.



\end{document}